\documentclass[aip,rsi,reprint, numerical]{revtex4-1}

\usepackage{graphicx}
\usepackage{epsfig}
\usepackage{lipsum}
\usepackage{color}
\definecolor{mygray}{gray}{0.6}
\definecolor{pink}{RGB}{255,200,200}
\definecolor{mygreen}{RGB}{0,150,0}

\begin{document}


\title{Simple, low-noise piezo driver with feed-forward for broad tuning of external cavity diode lasers} 



\author{S. Charles Doret}
\email[]{scd2@williams.edu}
\affiliation{Department of Physics, Williams College, Williamstown, MA  01267  USA}

\date{\today}

\begin{abstract}
We present an inexpensive, low-noise ($<260~\mu$V$_{rms}$, 0.1~Hz - 100~kHz) design for a piezo driver suitable for frequency tuning of external-cavity diode lasers.  This simple driver improves upon many commercially available drivers by incorporating circuitry to produce a `feed-forward' signal appropriate for making simultaneous adjustments to the piezo voltage and laser current, enabling dramatic improvements in mode-hop-free laser frequency tuning range.  We present the theory behind our driver's operation, characterize its output noise, and demonstrate its use in absorption spectroscopy on the rubidium D$_1$ line.   
\end{abstract}


\maketitle 

\section{Introduction}
\noindent External cavity diode lasers (ECDLs) have become pervasive in experiments throughout physics.  This is especially true in atomic, molecular, and optical physics, where the ECDL's broad spectral coverage, high efficiency, small size, and low cost have made them central to a wide array of experiments.\cite{WH91}  Excellent turn-key commercial laser systems are now available from a number of manufacturers, but these systems are typically many times more expensive than the sum of their constituent parts.  Fortunately, a number of researchers have developed low-noise current sources\cite{EVD08, TED11, SMC16e} and stable, robust opto-mechanical designs,\cite{AWB98, HWS01, BGB06e, CMB12e} thus it is quite feasible to build an inexpensive, research-grade laser and customize it to the needs of a particular experiment.  

A feature common to each of these ECDL designs is a movable optical element in the external cavity, the position of which may be adjusted to tune the laser's wavelength.   Fine position control over this element is typically achieved by tuning the voltage applied to a piezoelectric transducer (PZT).  Optimal laser performance thus requires the PZT be driven via an amplifier that produces moderately high voltages (typically $\sim$100~V) with very low noise, permitting laser frequency tuning over the PZT's full range without introducing spurious laser frequency jitter.  A recent paper\cite{PRR16e} describes an amplifier design utilizing active noise cancellation to achieve low-noise operation with outputs up to 250~V.  However, for many PZTs voltages of 75-100~V are adequate, permitting simpler circuits to achieve comparable low-noise performance.  Furthermore, it is particularly important that the amplifier have low noise in the low-frequency regime (below $\sim$10~kHz) where the motion of the ECDL grating is most responsive.  Finally, frequency tuning of an ECDL will typically be mode-hop-free for only a small range of frequencies unless an anti-reflection (AR) coated laser diode is used.  Unfortunately, AR coating of laser diodes requires specialized equipment, and such diodes are both less widely available and far more expensive than uncoated diodes.  It is thus desirable for the PZT driver to modulate the laser current in tandem with the PZT voltage to increase mode-hop-free tuning range.\cite{NFN98e,PLS01e}  

Here we discuss the design and performance of a simple piezo driver with several features satisfying these needs.  An on-board DC-DC converter generates the necessary high-voltage (HV) without the need for a specialized external supply.  Multiple input channels permit flexibility in external control of the amplifier output.   Thoughtful component selection and design choices that reduce AC-line noise contribute to exceptional low-noise performance at low frequencies where piezos are most commonly used in ECDLs.  Overall amplifier noise is comparable to or superior to that of a wide assortment of commercial drivers with published specifications, many of which cost an order of magitude more.   The circuit also includes a feed-forward element which produces a tunable low-voltage output which, when used to modulate the laser's current synchronously with the PZT, enables mode-hop-free tuning of the laser's output frequency over the full range of the piezo -- even when using laser diodes which are not AR coated.  The board and component cost is quite low ($\sim$ \$150), thus it is feasible to implement the design as needed throughout teaching and research laboratory settings.  Design schematics and related files are available online.\cite{PZT_site}

\section{Theory}
\subsection{ECDL Frequency Selection}
\begin{figure}[h]
\includegraphics[width=8.0 cm]{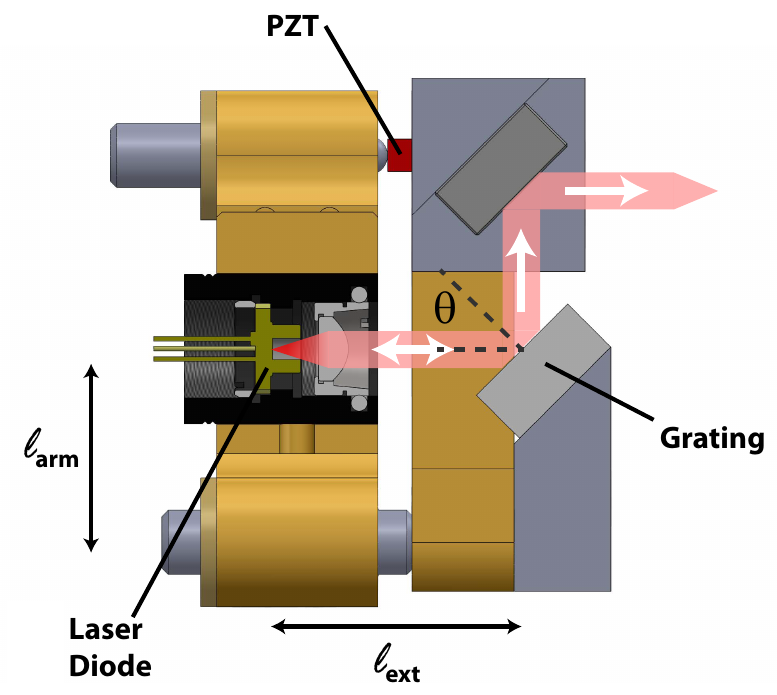}
\caption{A Littrow configuration external cavity diode laser in the style of [$\!\!$\citenum{HWS01}].}
\label{fig:Littrow}
\end{figure}

\begin{figure*}[]
\includegraphics[width=16 cm]{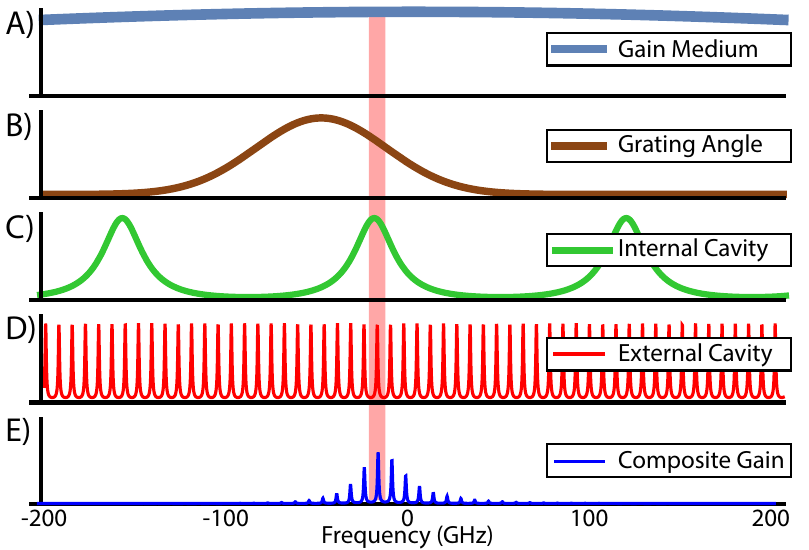}
\caption{(A-D) Contributions to ECDL laser frequency selection.  Here we assume the design from fig.~\ref{fig:Littrow} using a non-AR-coated 780~nm AlGaAs diode such that there are distinct internal and external cavities.  (E) Relative strengths of various longitudinal modes; mode competition ensures that the actual lasing mode will be the one with the largest overall gain, here highlighted by the pale red rectangle.}
\label{fig:mode_competition}
\end{figure*}

\noindent A number of different tunable ECDL designs have emerged, each featuring its own way of adjusting the laser's output frequency.  The most common of these is the Littrow configuration, in which the $m=-1$ order diffraction off of a reflectance grating is directed back into the laser diode to generate optical feedback (fig.~\ref{fig:Littrow}).  
Light striking the grating at an angle $\theta$ to the normal will diffract in a direction $\theta'$ such that
\begin{equation}
a \bigl( \sin \theta + \sin \theta' \bigr) = \lambda,
\end{equation}
where $a$ is the grating spacing and $\lambda$ the wavelength of the laser light.  To achieve optical feedback the diffracted light must propagate back along the direction of the incident light, further requiring that $\theta = \theta'$, so one can more simply write
\begin{equation}
\label{eqn:grating}
\sin \theta = \frac{\lambda}{2 a}.
\end{equation}
In addition to satisfying the grating angle criterion, laser light must also satisfy a constructive interference criterion for any optical cavities present in the laser.  
For an AR-coated laser diode there is just the `external' cavity, formed by reflection off of the back facet of the diode and diffraction off of the grating.  However, for a non-AR-coated diode there is also a second `internal' cavity, formed by reflections between the front and back facets of the diode (fig.~\ref{fig:mode_competition}c,d).\cite{thirdcavity}  As such, the combination of the grating angle and cavity interference effects gives rise to an array of competing modes that together define possible laser output frequencies; mode competition selects the mode with the largest overall gain (fig.~\ref{fig:mode_competition}e).    
 

Referring to equation~(\ref{eqn:grating}), it is straightforward to show that tuning the grating angle leads to a frequency shift of 
\begin{equation}
\frac{d f}{d \theta} = -\frac{2 a c}{\lambda^2} \cos \theta,
\end{equation}
($c$ is the speed of light).  Similarly, changing the length of an optical cavity by a length $\delta\ell$ will shift the resonant frequency of a particular longitudinal mode according to 
\begin{equation}
\frac{\delta f}{f} = -\frac{\delta \lambda_n}{\lambda_n}=-\frac{\delta \ell}{\ell}.
\end{equation}
Tuning the cavity length thus causes a frequency shift
\begin{equation}
\frac{df}{d\ell} = -\frac{c}{\lambda \ell}.
\end{equation}

\subsection{Single-mode tuning} 
\noindent Mode-hop-free tuning of the laser requires simultaneously adjusting the grating angle and cavity lengths so as to ensure that the same mode is associated with the largest gain at all times.  As an example, consider a 780~nm ECDL for which typical values of relevant parameters might be $a^{-1}=$~1800~lines/mm, $\ell_{ext} = 2$~cm, $\ell_{int}\approx 300$~$\mu$m, and refractive index $n_{int} = 3.7$.  In this case one finds
\begin{equation}
\label{eqn:tuning_rates}
\frac{df}{d\theta} = -390~\textrm{GHz/mrad} \hspace{.3in} \frac{df}{d\ell}_{ext} = -19~\textrm{GHz/$\mu$m}.
\end{equation}
For an AR-coated diode there is no internal cavity, thus mode-hop-free tuning can be achieved by setting $d\theta/d\ell_{ext}$ to satisfy equation~(\ref{eqn:tuning_rates}), i.e. 
\begin{equation}
\label{eqn:lever_arm}
\frac{d\theta}{d\ell} = .049~ \textrm{mrad/$\mu$m}.
\end{equation}
This merely requires the grating be mounted on a lever-arm of appropriate length (approximately 2~cm here; labeled as  $\ell_{arm}$ in fig.~\ref{fig:Littrow}), or by enabling independent control of angle and cavity length using synchronized control over two PZTs.\cite{HWS01}  In fact, choosing a grating line density $a$ such that $\theta \approx 45^\circ$ naturally gives rise to a lever arm that approximately satisfies equation~(\ref{eqn:lever_arm}) when the grating is centered on the faceplate of a standard one-inch optical mount, as in fig.~\ref{fig:Littrow}.  

With a non-AR-coated diode the situation is somewhat more complex.  As with an AR-coated diode, it is important that the grating angle and external cavity length be tuned in tandem.  However, now it is necessary to tune the positions of the internal cavity modes as well.  Otherwise the laser will hop between adjacent modes of the external cavity after being tuned in frequency by roughly half of the external cavity mode spacing - typically a few GHz.  Although the basic geometry of the internal cavity is fixed by the laser diode, it \emph{is} possible to tune the internal cavity using the laser diode injection current.  Current modulation leads both to thermal modulation of the diode's active region\cite{PD92} and modulation of $n_{diode}$;\cite{Nash73} thermal modulation dominates at frequencies below 10~MHz.  As such, by modifying the laser current in phase with variations in the PZT length one can shift internal cavity modes as needed to prevent mode-hops.  This is most easily achieved by modulating the laser current in tandem with the PZT drive voltage.\cite{NFN98e,PLS01e}  For a piezo installed as in fig.~\ref{fig:Littrow}, increasing PZT voltage (which lengthens the PZT) will lengthen the external cavity, decreasing the laser frequency.  Increasing the laser current will typically also decrease the frequency, so with the appropriate PZT voltage / laser current proportionality, mode-hop-free tuning is possible.

\begin{figure*}[t!]
\includegraphics[width=18.0 cm]{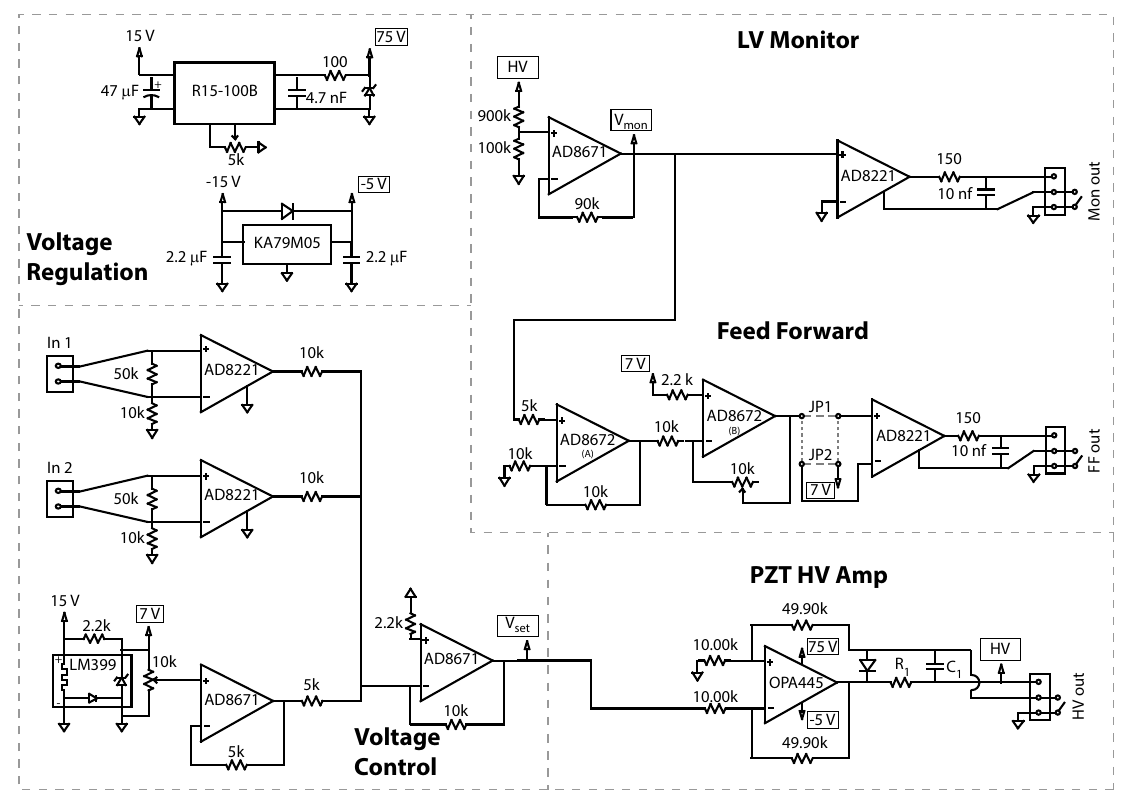}
\caption{Schematic of the PZT driver circuit; dashed lines subdivide the circuit for easy referencing to the text.}
\label{fig:circuit}
\end{figure*}
\section{PZT Driver Circuit Design}

\noindent Fig.~\ref{fig:circuit} breaks the circuit into several sections, each of which is described in further detail below.  Although not shown in the figure, each op-amp (excepting the high-voltage OPA445) is powered by $\pm$15 volts.  The circuit is fabricated using surface mount components on a 3$\times$4~in.$^2$ printed circuit board (PCB) with the board ground connected to the stamped aluminum enclosure in which it is mounted.  Surface-mount ceramic filter capacitors (0.1 and 10 $\mu$F) to ground are connected near the power pins of all op-amps to reduce RF noise.  With the exception of the resistor network for the HV amplifier, all of the resistors are standard 1/4 W, 200~ppm/$^\circ$C thick-film resistors in a 1206 surface-mount package.   All capacitors are 1206 surface-mount ceramic excepting the input filter capacitors for the on-board voltage sources which are solid tantalum.  

\subsection{Voltage Regulation}

\noindent In addition to the $\pm$15~V which powers most of the curcit elements, two additional voltages are needed to power the OPA445 HV amplifier.  A KA79M05 regulator reduces the -15V circuit supply to -5V to power the negative rail.  This leaves fully 95~V of the maximum 100~V rail-to-rail voltage tolerated by the OPA445 to be on the positive side, maximizing the positive swing of the amplifier output.  To source this voltage to the positive rail we utilize the RECOM R15-100B DC/DC converter, capable of sourcing 50~mA at up to 130~VDC.  A Zener diode at the R15-100B output eliminates output ripple and cleans up any noise at the switching frequency (200~kHz). 


\subsection{Voltage Control} 

\noindent AD8221 instrumentation amplifiers are used to generate high impedance analog inputs.  The AD8221s also decouple the inputs from the board ground to break any possible ground loops between control electronics and the amplifier circuit, eliminating this source of noise.  As drawn in fig.~\ref{fig:circuit} the AD8221 is set to unity gain; combined with the 5$\times$ gain of the HV amplifier this allows a standard $\pm10$~V DAC to swing the output through its full range.  However, the AD8221 gain is controlled with a single resistor, enabling straightforward modification for use with lower voltage DACs.  We include two identical analog inputs, suitable for locking applications where one input might be used for a feedback signal while the other supplied a high frequency dither, for example in stabilizing the DC length of a scanning Fabry-Perot cavity.  An additional voltage derived from a LM399 7~V reference controls a DC voltage offset.  For our application a traditional potentiometer offers adequate stability, but replacement with a digital setpoint (e.g. derived from an AD5541 DAC) would be a straightforward modification.    The analog input voltages are summed with the DC offset to produce the setpoint voltage sent to the high-voltage amplifier.  

\subsection{High-Voltage Amplifier}

\noindent The heart of the design is its high-voltage amplifier sub-circuit.  This amplifier is a standard differential amplifier designed around the OPA445, an inexpensive, low-noise op-amp capable of operation with up to 100~V between its rails.  We use asymmetric rail voltages of \mbox{-5~V} and up to +90~V (typically +75~V), appropriate for driving many standard piezoelectric transducers over essentially their full range.  A simple network of 0.01\% resistors sets the amplifier gain ($5\times$ in the figure).  A switch connects the output reference of the differential amplifier to ground.  For most applications the switch is closed, but by opening the switch the output can instead be referenced to an off-board ground.  This configuration helps to reduce line noise from possible ground loops for applications where the piezo is grounded at the ECDL.  A diode at the amplifier output prevents possible reverse biasing of any attached piezo, while an optional simple RC filter reduces high-frequency output noise for applications requiring only limited amplifier bandwidth. 

\subsection{Monitor and Feed Forward}

\noindent The output from the high voltage amplifier is resistively divided by ten and sent to two low-voltage output channels.  The first (LV Monitor) produces a faithful $\div$10 reproduction of the HV output suitable for monitoring on an oscilloscope or similar.  The second (Feed Forward) produces a proportional output with adjustable gain, shifted to be zeroed when the HV output is in the middle of its range.  In combination with jumpers (JP1, JP2) for switching the output polarity, the feed-forward signal can thus be adjusted as needed to optimize an ECDL's mode-hop-free tuning range.  The particular adjustment needed is highly dependent on the details of the external cavity, PZT, and laser diode, but a typical value might be $0.1$~mA/V (see section~\ref{sec:spectroscopy}).  As with the HV output, both low-voltage outputs can be switchably referenced to a distant ground, while 100~kHz low-pass filters eliminate spurrious RF noise. 

\section{Results}


\subsection{Amplifier Modulation Bandwidth and Noise}

\begin{figure}[h]
\includegraphics[width=8.8 cm]{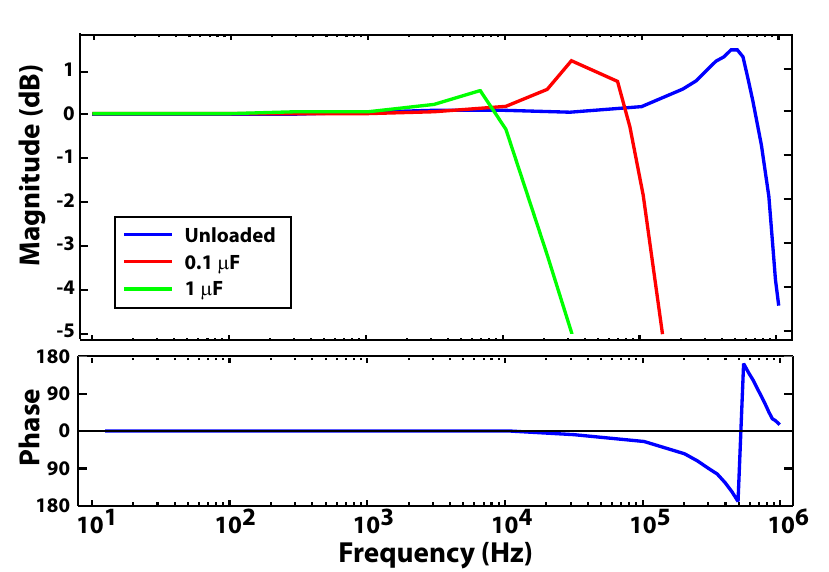}
\caption{(top) Small-signal modulation bandwidth of the piezo driver.  (bottom) Phase response for the unloaded driver.}
\label{fig:modulation}
\end{figure}

\noindent The large-signal (70~V$_{pp}$) bandwidth of the unloaded amplifier is roughly 50~kHz, limited by the OPA445 slew rate of 15~V/$\mu$s.   The small-signal bandwidth (fig.~\ref{fig:modulation}) for the unloaded amplifier is approximately 400~kHz, as expected from the 2~MHz gain bandwidth product of the OPA445 given the 5$\times$ gain of the HV differential amplifier.  The relatively modest output current capacity of the OPA445 (26~mA short-circuit current) limits the driver bandwidth to 10~kHz (100~kHz) for 1~$\mu$F (0.1~$\mu$F) capacitive loads.      

\begin{figure}[]
\includegraphics[width=8.8 cm]{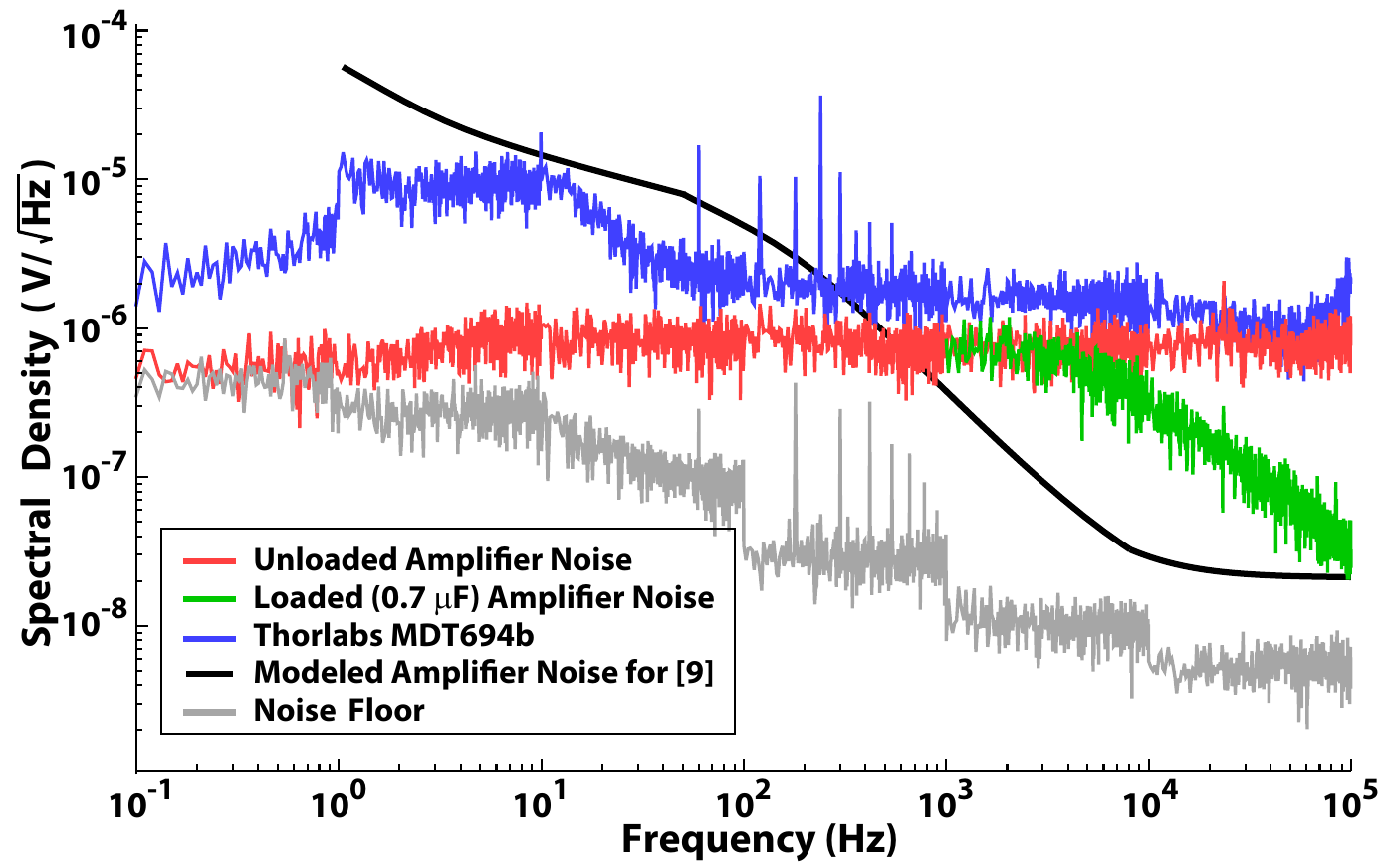}	
\caption{PZT driver voltage noise spectral density.    \textcolor{red}{\textbf{(red)}} noise spectrum of the PZT driver described here with no load applied, measured with the output at 5~VDC.  The superimposed high frequency tail shown in \textcolor{mygreen}{\textbf{(green)}} shows the noise when a 0.7~$\mu$F capacitor is connected across the output, typical of the capacitance of a PZT.  Two additional noise spectra are included for comparison: a common commercial PZT driver (Thorlabs MDT694b) \textcolor{blue}{\textbf{(blue)}}, also at 5~VDC, and the noise model presented for the design by Pisenti et al.\cite{PRR16e} \textcolor{black}{\textbf{(black)}}.  The noise floor for the measuremed traces is plotted in \textcolor{mygray}{\textbf{(gray)}}.}
\label{fig:noise}
\end{figure}

In fig.~\ref{fig:noise} we plot the noise power spectral density of our amplifier. 
 These data were acquired using a Stanford Research SR560 low-noise voltage preamplifier (AC coupled, 1000x gain) and a Tektronix TDS3014B oscilloscope.  The total integrated noise (0.1~Hz - 100~kHz) for the unloaded amplifier is 260~$\mu$V$_{rms}$.  Of particular note is the superb performance in the low-frequency regime; the total low-frequency noise (0.1~Hz - 10~Hz) for our driver is only 2.6~$\mu$V$_{rms}$.  Many commercial PZT drivers fail to account for this low-frequency noise when quoting performance, instead specifying RMS noise as measured by a multimeter (20~Hz to 100~kHz) - often a significant underestimate.   Further, this low-frequency output noise is particularly troublesome; such low-frequency noise can be eliminated only through active, closed-loop feedback and is \emph{not} reduced in the presence of a capacitive load such as a PZT.  In contrast, the addition of a capacitive load to our driver (0.7~$\mu$F, characteristic of a typical PZT),\cite{resistor} reduces the total integrated noise to 68~$\mu$V$_{rms}$ (fig.~\ref{fig:noise}).  Significant further reductions are possible for low frequency applications, for example to $\sim 10~\mu$V$_{rms}$ with a 100~Hz bandwidth.

\subsection{Mode-hop-free Spectroscopy}
\label{sec:spectroscopy}
\begin{figure}[h]
\includegraphics[width=8.8 cm]{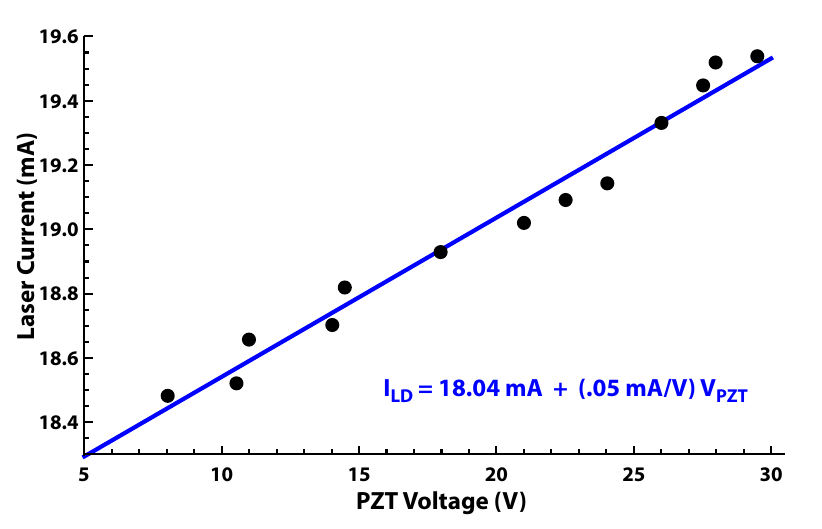}
\caption{Empirical determination of feed-forward proportionality.  For the laser described here the appropriate proportionality constant is $\sim$0.05~mA / V.}
\label{fig:proportionality}
\end{figure}

\noindent Mode-hop-free tuning is accomplished by modulating the laser current in tandem with the PZT voltage so as to keep the internal and external cavity mode frequencies well aligned.  The appropriate proportionality is dependent on the details of the PZT, external cavity geometry, and laser diode, but may be easily determined empirically.  To do so, one can slowly tune the PZT voltage in the absence of any current modulation while watching for laser mode hops.  The laser current can then be minutely adjusted to return the laser to the original mode.   Detecting mode hops and correcting for them with adjustments to the laser current is most easily achieved using a wavemeter to monitor the laser's output frequency.  It is also straightforward to detect hops by monitoring the laser output with a scanning Fabry-Perot etalon or even by observing discrete jumps in the laser's output power on a photodiode.  Detecting/correcting mode hops over a range of PZT voltages and plotting the relationship between PZT voltage and laser current yields the necessary proportionality constant, as in fig.~\ref{fig:proportionality}.  

\begin{figure}[]
\includegraphics[width=8.8 cm]{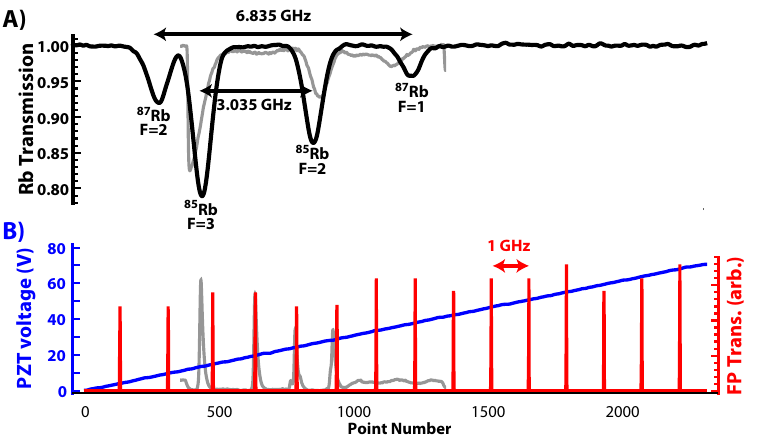}
\caption{Absorption spectroscopy of a rubidium vapor cell.  The heavy black curve in (A) shows a mode-hop-free scan over essentially the full range of the PZT driver (roughly 15~GHz for the laser used here), showing the familiar D$1$ lines.  (B) Voltage \textcolor{blue}{\textbf{(blue)}} applied to the laser PZT, and transmission \textcolor{red}{\textbf{(red)}} through a confocal Fabry-Perot etalon (FSR = 1 GHz).  The same scan performed with the feed-forward signal to the laser disconnected is shown on both plots in gray.  Mode hops can be seen as discontinuities in the vapor cell  transmission in (A), while the corresponding trace of Fabry-Perot transmission in (B) clearly shows a non-linear relationship between PZT voltage and laser frequency as well as regions of multi-mode behavior.}
\label{fig:spectroscopy}
\end{figure}

To determine the efficacy of our piezo driver's feed-forward element and demonstrate mode-hop-free laser-frequency tuning we performed absorption spectroscopy on the rubidium D1 lines.  Transmission through a Fabry-Perot etalon (1~GHz free spectral range) provides additional frequency references and confirms the linearity of the scan.   Using an ECDL of the type shown in fig.~\ref{fig:Littrow} with a 780~nm laser diode (Thorlabs L780P010), a small piezoelectric actuator (Thorlabs AE0203D04F, 4.6 $\mu$m stroke @ 150~V) scans the ECDL grating angle, while the feed-forward circuit is adjusted to drive the external modulation input on a commercial current driver (Thorlabs LDC201CU).  With the feed-forward engaged, the ECDL could be scanned without mode-hops over the full range of the driver (fig.~\ref{fig:spectroscopy}).  This equates to $\sim$15~GHz for our laser, versus $\sim$3~GHz with the feed-forward disengaged.  Still larger tuning ranges can be achieved by using a piezo with a larger stroke or by driving the grating with a piezo mounted closer to the pivot of the grating arm.

\section{Summary}

\noindent We have developed a simple piezoelectric driver tailored for use with external cavity diode lasers.  The design features a differential amplifier based around an inexpensive, low-noise op-amp and an on-board DC-DC converter, making it easy to understand and implement and economical to deploy.  The design achieves low-noise performance over a wide bandwidth, and is exceptionally quiet in the low-frequency regime where grating-stabilized lasers are most sensitive to PZT voltage noise.  In addition to the high-voltage output, the design also generates a tunable low-voltage feed-forward signal, permitting extended mode-hop-free laser frequency tuning to facilitate spectroscopy and make laser frequency locks more robust. 

\vspace*{.3in}

%
%

%

\begin{acknowledgments}
\noindent Thanks to Tiku Majumder for helpful comments on the manuscript, and to students Iona Binnie, Derek Galvin, and Hallee Wong who assembled early versions of this design.  This work has been supported by a Cottrell College Science Award from the Research Corporation for Science Advancement and by Williams College.  
\end{acknowledgments}

\bibliographystyle{aipnum4-1}

\bibliography{PZT_RSI_bib}

\begin{thebibliography}{16}
\bibitem [{thi()}]{thirdcavity}%
  \BibitemOpen
  \href@noop {} {}\bibinfo {note} {In principle there is also a third cavity
  formed by the diode's front facet and the diffraction grating. However, the
  finesse of this cavity is so low as to play essentially no role in mode
  selection.}\BibitemShut {Stop}%
\bibitem [{res()}]{resistor}%
  \BibitemOpen
  \href@noop {} {}\bibinfo {note} {We have also added a 47~$\Omega$ resistor in
  series with the capacitor to form a low-pass RC filter with a well-defined
  frequency corner of 5~kHz.}\BibitemShut {Stop}%
\end{thebibliography}

\end{document}